\input{aipcheck}

\documentclass[
    ,final            
  ]
  {aipproc}
  \pdfoutput=1
  
  \usepackage{floatflt}
  \usepackage{deluxetable}
\def\lesssim{\mathrel{\hbox{\rlap{\hbox{\lower4pt\hbox{$\sim$}}}\hbox{$<$}}}}
\def\gtrsim{\mathrel{\hbox{\rlap{\hbox{\lower4pt\hbox{$\sim$}}}\hbox{$>$}}}}

\layoutstyle{6x9}

\begin{document}

\newcommand{\edot}{\dot{E}}
\newcommand{\pdot}{\dot{P}}
\newcommand{\lpwn}{L_{\rm pwn}}
\newcommand{\lpsr}{L_{\rm psr}}
\newcommand{\etapwn}{\eta_{\rm pwn}}
\newcommand{\etapsr}{\eta_{\rm psr}}
\newcommand{\chan}{{\sl Chandra}\/}
\newcommand{\be}{\begin{equation}}
\newcommand{\ee}{\end{equation}}
\newcommand{\xmm}{{\sl XMM-Newton\/}}
\newcommand{\suz}{{\sl Suzaku\/}}
\newcommand{\fermi}{{\sl Fermi\/}}

\title{Pulsar-wind nebulae in X-rays and TeV $\gamma$-rays}


\author{O. Kargaltsev}{
  address={University  of Florida, Bryant Space Center, Gainesville,
FL 32611, USA}
}

\author{G. G. Pavlov}{
  address={Pennsylvania State University, 525 Davey Lab., University Park,
PA 16802, USA}
}

\begin{abstract}
 Pulsars  are known to be efficient accelerators that produce copious amounts of relativistic 
 particles and inject them into the Galactic medium.  The radiation emitted by  such a pulsar wind can be seen from radio through $\gamma$-rays as a pulsar-wind nebula (PWN).
  Here we overview and summarize recent progress in X-ray and TeV observations of PWNe. 
\end{abstract}

\maketitle


 Along with supernovae and microquasars,   pulsars 
  are 
among the primary sources of the leptonic cosmic rays. 
  The energies of
  the pulsar wind 
  electrons and positrons 
range from 
 $\lesssim$1 GeV  to  
 $\sim$1 PeV,
   placing their synchrotron and inverse Compton (IC) emission into radio-X-ray and GeV-TeV bands, respectively.
 This multiwavelength emission can be seen as a {\em pulsar-wind nebula} 
(PWN; see \cite{2006ARA&A..44...17G} and \cite{2008AIPC..983..171K} for reviews).
  During the last decade,  observations with modern X-ray and TeV  observatories  have dramatically increased the number of known PWNe whose properties are summarized in this review.

\begin{table}
 \setlength{\tabcolsep}{0.05in}
 \begin{tabular}{rlcccccc}
\hline
 \tablehead{1}{c}{b}{   \# \\  }  &
  \tablehead{1}{c}{b}{  PSR  \\  } &
    \tablehead{1}{c}{b}{  PWN$^*$
 \\  } &
       \tablehead{1}{c}{b}{  VHE src.$^\dag$
\\  } &
      \tablehead{1}{c}{b}{  $\log\dot{E}$ \\ ${\rm [erg/s]}$  } &
      \tablehead{1}{c}{b}{  $\log\tau$  \\ ${\rm [yr]}$  } &
   \tablehead{1}{c}{b}{  $d^{**}$
\\  ${\rm kpc}$ } &

    \tablehead{1}{c}{b}{  ${\rm Rad./H}_{\alpha}/{\rm GeV}^\ddag$
\\  }
      \\
\hline
1   &   J0537--6910             &   N157B            & ...              &   38.68          &   3.70    &     50~~~~~      &  Y/N/N     \\
2   &   B0531+21                   & Crab            &  H~J0534+220     &   38.66          &   3.09   &       2~~~        &  Y/N/Y      \\
3    &    J2022+3842              & G76.9+1.0       & ...               &   38.30          &   3.95    &     8~~~          &  Y/N/N    \\
4   &   B0540--69                 &    N158A           & ...            &   38.17         &   3.22   &       50~~~~~       &  Y/N/N     \\
5   &   J1813--1749             & G12.82--0.02     & H~J1813--178$^?$   &  37.83        &   3.66     &       4.5       &  N/N/P \\
6     &   J1400--6325             &   G310.6--1.6     & ...              & 37.71           &  4.10   &       6~~~~          &    Y/N/N  \\
7   &   J1833--1034           &   G21.50--0.89      & H~J1833--105       &  37.52      &   3.69    &    4.8              &  Y/N/Y      \\
8   &   J0205+6449            &   3C\,58              & ...             &   37.43     &     3.73    &    3.2              &  Y/N/Y      \\
9   &   J2229+6114          &  G106.65+2.96   & V~J2228+609       & 37.35  &   4.02   &      3~~~                &  Y/N/Y     \\
10   &   B1509--58              &   Jellyfish             &  H~J1514--591       &   37.25     &   3.19  &    5~~~               &  P/N/N      \\
11   &   J1617--5055           &  G332.50--0.28  & H~J1616-508$^?$   & 37.20      &   3.91  &     6.5             &  N/N/P      \\
12   &   J1124--5916           &   G292.04+1.75   & ...                          &  37.07   &     3.46   &      6~~~               &  Y/N/Y     \\ 
13  &   J1930+1852           &  G54.10+0.27     & V~J1930+188      &   37.06   &   3.46   &    6.2~~             &  Y/N/N      \\
14  &   J1420--6048            &  G313.54+0.23  &  H~J1420--607    &   37.02   &   4.11   &      5.6              &  P/N/N     \\
 15  &   J1846--0258             &  Kes\,75           &   H~J1846--029     &  36.91   &   2.86    &     6$^{??}$~~                  &  Y/N/N      \\
16  &   B0833--45                & Vela                 &   H~J0835--455    &  36.84     &   4.05   &   ~~~~$0.29^{\rm p}$    &  Y/N/N      \\
17  &   J1811--1925             &   G11.18--0.35  & ...                             &   36.81    &   4.37   &    5~~~            &  Y/N/N      \\
18     &   J1838--0655              &     G25.24--0.19 &  H~J1837--069     &  36.74    &   4.36   &      7~~~                 &  N/N/N \\
19     &   J1418--6058             &  Rabbit               &  H~J1418--609    &    36.69   &   4.00    &           3$^?$~~              & Y/N/Y   \\
20     &   J1856+0245             &   G36.01+0.06$^?$          &  H~J1857+026  & 36.66      & 4.32     &      9~~~               & N/N/N \\
21  &   B1951+32                &  G68.77+2.82       &  ...                       &  36.57    &   5.03   &    2.5              &  Y/Y/Y      \\
22  &   J1826--1256           & Eel                                       &  ...                   &  36.55         &  4.15   &      7$^?$~~                &  P/N/Y   \\
23  &   J2021+3651           &   G75.23+0.12   &   M~J2019+37$^?$     &36.53   &   4.23   &    4~~~            &  N/N/Y    \\
24  &   B1706--44               &   G343.10--2.69     & H~J1708--443 &   36.53   &   4.24   &     2~~~            &  Y/N/Y     \\
25  &   J1357--6429           &  G309.92--2.51    & H~J1357--645  &   36.49   &   3.86    &    2.5              &  P/N/N      \\
26     &  J1913+1011            &   G44.48--0.17$^?$       & H~J1912+101  &  36.46   &   5.23     &    4.5              & N/N/N   \\
27     &   J1907+0602           &  G40.16--0.89$^?$        &  H~J1908+063$^?$ & 36.45   &  4.28     &       3.2             & N/N/Y  \\
28  &   B1823--13              &  G18.00--0.69      &  H~J1825--137  &  36.45   &   4.33   &      4~~~            &  N/N/N     \\
29  &   B1757--24             &            Duck                & ...                     &   36.41   &   4.19   &          5~~~  &  Y/N/N     \\
30  &   J1016--5857          &   G284.08--1.88    &  ...                       &   36.41   &   4.32  &        3~~~       &  N/N/N      \\
31  &   J1747--2958          &             Mouse          & ...                     &   36.40        &   4.41   &        5~~~   &  Y/N/Y     \\
32  &   J1119--6127           &    G292.15--0.54   & H~J1119--615$^?$   & 36.37   &   3.21   &   8.4             &  N/N/N      \\
33  &   B1800--21              &  G8.40+0.15            &  H~J1804--216$^?$ & 36.34     &   4.20   &    4~~~            &  N/N/P      \\
34  &   B1046--58            &  G287.42+0.58      &   ...                         & 36.30           &   4.31      &   3~~~            &  N/N/Y     \\
35  &   J1809--1917          &  G11.09+0.08       &   H~J1809--193        &  36.25  &   4.71    &    3.5              &  P/N/N       \\
36  &   J1301--6305           &  G304.10--0.24     & H~J1303--631$^?$  &  36.22   &   4.04   &    7~~~                &  N/N/N       \\
37  &   J1718--3825           &  G348.95--0.43    & H~J1718--385$^?$  &   36.11   &   4.95    &    4~~~             &  N/N/Y       \\
38     & J1531-5610               &   G323.89+0.03    &   ...                    &   35.96  &   4.99    &    3~~~            &  N/N/N    \\
39  &   J1509--5850           &  G319.97--0.62     & ...                       &   35.71  &   5.19    &    4~~~           &  P/N/Y       \\
40     &  J1857+0143            &       G35.17--0.57$^?$      & H~J1858+020$^?$  & 35.65    &   4.85    &   5.5        & N/N/N  \\
41     &   J0007+7303           &   CTA1                    & ...                       &  35.65   &    4.15   &        1.4               &  N/N/Y   \\
42  &   B1853+01              &   G34.56-0.50      &   ...                        & 35.63      &   4.31   &   3~~~        &  Y/N/N       \\
43     &  J1809-2332             &                                     Taz      & ...                          &    35.60    &    4.36  &  2~~~~      &  Y/N/N      \\
44     &   J1958+2846           & G65.89--0.37$^?$      &   M~J1954+28       &  35.58    &   4.32  &   2$^?$~~       & N/N/Y  \\
45  &   J1702--4128          &  G344.74+0.12      &  H~J1702--420$^?$ &  35.53    &   4.74   &    5~~~        &  N/N/N      \\
46  &   J0729--1448          &  G230.39--1.42     &  ...                                      &  35.45   &     4.54   &     4~~~      &  N/N/N       \\
47     &   J2032+4127           &     G80.22+1.02    &         He~J2032+4130          &  35.43   &     5.04   &     1.7          &  N/N/Y     \\
48  &   J1740+1000            &  G34.01+20.27    & ...                                       &   35.36   &   5.06    &    1.4              &  N/N/N       \\
        \hline
\end{tabular}
\caption{Pulsars with X-ray and/or TeV PWNe}
\label{tab:prop1}
\end{table}
\normalsize

\addtocounter{table}{-1}
\begin{table}
 \setlength{\tabcolsep}{0.05in}
 \begin{tabular}{rlcccccc}
\hline
 \tablehead{1}{c}{b}{   \# \\  }  &
  \tablehead{1}{c}{b}{  PSR  \\  } &
    \tablehead{1}{c}{b}{  PWN\tablenote{PWN name or galactic coordinates. The superscript $^?$ mark the cases in which no X-ray PWN has been reported, but there are
TeV PWN candidates nearby.  } \\  } &
       \tablehead{1}{c}{b}{  VHE src.\tablenote{TeV sources in the vicinity
of the PSR/PWN. `H', `V', `M' and `He' stand for HESS, VERITAS, Milagro and HEGRA.
 The superscript $^?$ marks questionable associations.} \\  } &
      \tablehead{1}{c}{b}{  $\log\dot{E}$ \\ ${\rm [erg/s]}$  } &
      \tablehead{1}{c}{b}{  $\log\tau$  \\ ${\rm [yr]}$  } &
   \tablehead{1}{c}{b}{  $d$\tablenote{Our best guess for the pulsar distance, used to scale the distance-dependent parameters in Table 2. The superscript $^{\rm p}$ marks the distances determined from parallax measurements; the most uncertain distances (e.g., when even dispersion measure is unknown) are marked by $^?$. For Kes 75 the distance (marked by $^{??}$) is rather uncertain, ranging from 5.1--7.5 kpc (\cite{2008A&A...480L..25L}) to $\sim10.6$ kpc (\cite{2009ApJ...694..376S}).  } \\  ${\rm kpc}$ } &
    \tablehead{1}{c}{b}{  ${\rm Rad./H}_{\alpha}/{\rm GeV}$\tablenote{Is the PWN detected in radio/H$_{\alpha}$, and PSR/PWN in GeV $\gamma$-rays? P = `possibly'.  \vspace{-0.9cm} } \\  } 
      \\
\hline
49     &  J0631+1036            &  G201.22+0.45$^?$       &   M~J0630+10$^?$  &   35.24  &   4.64    &    3.6             & N/N/Y \\
50  &   B1957+20               &  G59.20--4.70      &  ...                                       &  35.20    &   9.18    &    2.5              &  N/Y/N      \\
51     &   J0633+0632           &  G205.10--0.93         &    ...                               &  35.08   &  4.77     &    ~~1.5$^?$              &  N/N/Y \\
52     &  J1740-3015              &  G358.29+0.24$^?$                &  H~J1741--302               &  34.91   &   4.31    &    3~~~                 &   N/N/N  \\
53  &   J0538+2817            &  G179.72--1.69     &  ...                                   &  34.69  &   5.79   &   ~~~~$1.47^{\rm p}$    &  N/N/N      \\
54  &   B0355+54                 &       Mushroom     & ...                                      &   34.66      &   5.75   &   ~~~~$1.04^{\rm p}$    &  N/N/N       \\
55  &   J0633+1746             &      Geminga       & M~J0632+17$^?$      &   34.51     &   5.53   &     ~~~~$0.25^{\rm p}$   &  N/N/Y       \\
56     &   J1745--3040              &   G358.55--0.96$^?$                  &   H~1745--305$^?$        &  33.93      &   5.74   &       2~~~     & N/N/N \\
57     & J1502--5828               & G319.39+0.13$^?$         &  H~J1503--582$^?$          &   33.68    &   5.46      &     8~~~                       &     N/N/N  \\          
58  &   B1929+10                &  G47.38--3.88  &      ...                                      &   33.59   &   6.49      &  ~~~~$0.36^{\rm p}$   &  P/N/N     \\
59  &   B2224+65              &          Guitar            & ...                                      &   33.07   &   6.05      &     1.5       &  N/Y/N      \\ 
 \hline
\end{tabular}
\caption{Pulsars with X-ray and/or TeV PWNe (continued)}
\label{tab:prop1}
\end{table}

\begin{table}
 \setlength{\tabcolsep}{0.045in}
 \begin{tabular}{rccccclllc}
\hline
 \tablehead{1}{c}{b}{   \# \\  }  &
  \tablehead{1}{c}{b}{   $N_{\rm H,22}$$^\ast$ \\  }  &
   \tablehead{1}{c}{b}{   $\log L_{\rm X}$$^\dagger$ \\ ${\rm [erg/s]}$ }  &
   \tablehead{1}{c}{b}{   $\Gamma_{\rm X}$  \\  }  &
   \tablehead{1}{c}{b}{   $\log L_\gamma$$^{\ast\ast}$ 
\\ ${\rm [erg/s]}$  }  &
    \tablehead{1}{c}{b}{   $\Gamma_{\gamma}$$^\ddagger$
 \\  }  &
      \tablehead{1}{c}{b}{   $l_{X}$$^\S$ 
\\  ${\rm pc}$ }  &
  \tablehead{1}{c}{b}{   $l_{\gamma}$$^\P$ 
\\ ${\rm pc}$ }  &
 \tablehead{1}{c}{b}{   $\Delta$$^\parallel$
 \\ ${\rm arcmin}$ } &
\tablehead{1}{c}{b}{ ${\rm Refs.}$$^{\dagger\dagger}$ 
\\}
  \\
\hline
1       &     0.5   &    $36.04  \pm   0.01$    &     $2.20\pm0.05$       &   ...                            &      ...                            &  1.4  &    ...       &      ...     &  \cite{2006ApJ...651..237C}    \\
2       &     0.32  &    $37.28  \pm   0.01$    &     $2.12\pm0.01$       &   $34.51  \pm  0.06$   &      $2.63\pm0.20$          &  1.2  &   $  <2  $ &   $  <0.5   $    &  \cite{2004ApJ...609..186M, 2001AnA...365L.212W, 2006AnA...457..899A}    \\
3        &     1.3    &    $32.58  \pm  0.12$    &      $0.9\pm0.5$          &   ...                            &    ...                                &  0.4  &    ...     &     ...       &   \cite{2009_Arzoumanian_Boston}  \\
4       &     0.46  &    $37.01  \pm   0.01$    &     $1.85\pm0.10$       &   ...                          &      ...                             &  1.4  &   ...         &   ...    &  \cite{2001ApJ...546.1159K}    \\
5       &     10~~~~~~~    &    $32.90\pm0.25$        &     $0.4^{+0.4}_{-0.7}$ &  $34.34 \pm 0.14$   &     $2.09\pm0.21$            &  2.0  &    $  6  $   &   $< 0.5$         & \cite{2007ApJ...665.1297H, 2006ApJ...636..777A,	2009ApJ...700L.158G}\\  
6        &     2.1    &    $ 34.99\pm0.04$       &     $1.83\pm0.09$       &   ...                           &   ...                              &  2.3  &  ...             &  ...                  &  \cite{2009arXiv0910.3074R} \\  
7       &     2.3   &    $35.36  \pm   0.01$    &     $1.89\pm0.02$       &   $33.63 \pm 0.12$    &      $2.08\pm0.22$          &  1.0  &   $<5$     &   $ < 1 $    &  \cite{2001ApJ...561..308S, 	2007arXiv0710.2247H}   \\
8       &     0.43  &    $33.94  \pm   0.01$    &     $2.02\pm0.01$       &   ...                          &    ...                              &  1.2 &   ...          &    ...            &  \cite{2004ApJ...616..403S}    \\
9       &     0.5   &    $32.94  \pm   0.01$    &     $1.3\pm0.1$         &   $33.63 \pm 0.20$     &      $2.3\pm0.4$             &  0.4  &   $ 26   $ &   $  24   $    &  \cite{2001ApJ...552L.125H, 2009_Aliu_Boston}    \\
10       &     0.8   &    $34.60  \pm   0.03$    &     $1.65\pm0.05$       &   $34.86  \pm  0.12$   &      $2.27\pm0.2$            &  4.5  &   $  24  $ &   $   2.4 $    &  \cite{2002ApJ...569..878G, 2005AnA...435L..17A}  \\
11       &     3.5   &    $33.79  \pm   0.02$    &     $1.2\pm0.2$         &   $34.78  \pm  0.06$   &      $2.35\pm0.21$          &  0.6  &   $60 $ &   $   10  $    & \cite{2009ApJ...690..891K, 2006ApJ...636..777A}     \\
12       &     0.37  &    $34.71  \pm   0.03$    &     $1.7\pm0.5$         &   ...                           &     ...                             &  0.2  & ...     &   ...    & \cite{2001ApJ...559L.153H} \\
13      &     1.9   &    $34.46  \pm   0.01$    &     $1.99\pm0.03$       &   $33.88  \pm  0.20$   &      $2.3\pm0.4$          &  1.5  &   $  <11 $ &   $   <3 $    & \cite{2002ApJ...568L..49L, 2009_Aliu_Boston}  \\
14      &     5.4   &    $33.15  \pm   0.11$    &     $0.5\pm 1.2$         &   $34.60  \pm  0.07$   &      $2.17\pm0.12$        &  0.4  &   $11$   &   $3.3$    &  \cite{2005ApJ...627..904N, 2006AnA...456..245A} \\
15      &     4.0   &    $35.19  \pm   0.02$    &     $2.03\pm0.02$       &   $33.87  \pm  0.09$   &      $2.26\pm0.15$          &  2.8  &   $<6 $ &   $<1$    & \cite{2003ApJ...582..783H}  \\
16      &     0.02  &    $32.11  \pm   0.03$    &     $1.4\pm0.1$         &   $32.14  \pm  0.22$   &      $1.45\pm0.22^?$  &  0.1  &   $ 5 $ &   $  30   $    &    \cite{2001ApJ...554L.189P, 2001ApJ...556..380H, 2006AnA...448L..43A} \\
17      &     3.1   &    $34.00  \pm   0.09$    &     $1.5\pm0.2$         &  ...                           &      ...                             &  1.0  &   ...        &  ...    & \cite{2003ApJ...588..992R}  \\
18         &      4.0   &    $33.47\pm0.14$        &      $0.8\pm0.3$          &  $34.91 \pm 0.18$       &      $2.27\pm0.21$           &   2.0  &  28       &    3    &\cite{2008ApJ...681..515G, 2006ApJ...636..777A} \\
19       &      2.3   &    $33.55\pm0.02$        &    $1.7\pm0.1$            &  $33.91\pm0.07$       &     $2.22\pm0.12$            &  1.5 &   8         &    3.5  &   \cite{2005ApJ...627..904N, 2006AnA...456..245A}  \\
20       &  $\lesssim2~~~$  &    ...                     &      ...                         &  $35.16\pm0.11$       &     $2.39\pm0.08$           &  1.7  &  40       &    6     & \cite{ 2008ApJ...682L..41H, 2008AnA...477..353A}\\  
21      &     0.34  &    $32.62  \pm   0.01$    &     $1.76\pm0.03$       &   ...                          &      ...                              &  0.4 &   ... &   ...    &  \cite{2004ApJ...610L..33M, 2005ApJ...628..931L}  \\
22        &       2.0  &   $32.41 \pm 0.05$       &     $1.27\pm0.40$      &  ...                            &      ...                               & 0.3   & ...   & ...   &  \cite{2007AAS...21114403R} \\
23      &     0.7   &    $33.08  \pm   0.07$    &     $1.7\pm0.3$         &   $33.09\pm0.50$       &       
                                                                                                       [2.6]   &  0.8  &   $ <140 $ &   $  18  $    & \cite{2004ApJ...612..389H, 2009ApJ...700L.127A} \\
24      &     0.5   &    $32.58  \pm   0.02$    &     $1.8\pm0.1$         &   $33.87  \pm  0.13$   &      $2.00\pm0.22$            &  0.2  &   $ 23 $ &   $   14  $    & \cite{2005ApJ...631..480R, 2009arXiv0906.5574H}  \\
25      &    0.23   &    $31.70   \pm  0.07$    &     $1.28\pm0.21$     &   $33.77   \pm 0.20$   &      $2.2\pm0.2$            &  0.1 &   $20$ &   $8   $    & \cite{2007ApJ...665L.143Z}   \\
26       & $\lesssim2$   &     $<31.22$           &    ...                         &  $34.19  \pm  0.19$    &      $2.7\pm3.6$            & ...      &   $60$ &   $12$     & \cite{2008AnA...484..435A} \\
27       & $\lesssim2$   &     ...                      &     ...                         &  $34.22 \pm 0.22$      &      $2.10\pm0.21$        & ...      &   $37$ &   $16$     & \cite{2009AnA...499..723A} \\        
28      &     1.0   &    $32.50  \pm   0.05$    &     $1.3\pm0.4$         &   $35.05  \pm  0.10$   &      $2.26\pm0.2^?$  &  0.2  &   $70$ &   $10$    &  \cite{2003ApJ...588..441G, 2008ApJ...675..683P, 2006AnA...460..365A}  \\
29      &     4.4   &    $33.20  \pm   0.14$    &     $2.5\pm0.3$         &   ...                            &     ...                             &  0.5  &   ...    &   ...   &  \cite{2001ApJ...562L.163K}   \\
30      &  $\!\!\!$[1.2] &    $32.30  \pm   0.11$    &     $1.5\pm0.2$  &   ...                           &     ...                             &  0.1  & ...      &   ...     &  \cite{2004ApJ...616.1118C}   \\
31      &     3.0   &    $34.70  \pm   0.05$    &     $2.0\pm0.2$         &  ...                           &     ...                              &  0.5  &  ...     &   ...     &  \cite{2004ApJ...616..383G}   \\
32      &     1.6   &    $33.00  \pm   0.10$    &     $1.5\pm0.3$         &   $34.20  \pm  0.30$   &      $>2.2$                     &  0.5  &  $30$ &   $6$    &  \cite{2003ApJ...591L.143G, 2009_Djannati_Boston}   \\
33      &     1.4   &    $32.20  \pm   0.05$    &     $1.6\pm0.3$         &   $34.30  \pm  0.08$   &      $2.72\pm0.21$          &  0.2  &   $  58 $ &   $   10  $    &  \cite{2007ApJ...660.1413K, 2007ApJ...670..643K, 2006ApJ...636..777A}  \\
34      &  $\!\!\!$[0.4] &    $31.82  \pm   0.04$    &     $1.0\pm0.2$  &   ...                           &     ...                               &  0.2  &   ...    &   ...    &  \cite{2006ApJ...652..569G}   \\
35      &     0.7   &    $32.59  \pm   0.03$    &     $1.4\pm0.1$         &   $34.29  \pm  0.11$   &      $2.20\pm0.22$           &  0.2 &   $40$  &   $8$    &  \cite{2007ApJ...670..655K,  2007AnA...472..489A}    \\
36      &  $\!\!\!$[1.1] &    $32.16   \pm  0.50$    &     ...                 &   $34.76   \pm 0.08$   &      $2.44\pm0.21$           &  2.0  &  $39$  &   $11$      &  \cite{2005AnA...439.1013A}  \\
37      &     0.7   &    $32.60   \pm  0.10$    &     $1.9\pm0.2$         &   $33.74   \pm 0.30$   &      $0.7\pm0.6^?$   &  2.0    &  $7$    &     3       &  \cite{2007AnA...476L..25H, 2007AnA...472..489A}    \\
38        &  $\lesssim2$  &    $31.03 \pm  0.5$   &     ...                    &  ...                             &       ...                             &
                                                                                                                                                           
                                                                                                                                      0.1  & ...        &   ...         & ...            \\
39      &     2.1   &    $32.12  \pm   0.13$    &     $1.8\pm0.3$         &   ...                            &      ...                             &  0.4  &  ...        &   ...          &  \cite{ 2008ApJ...684..542K}    \\
40       &   $\lesssim2$  &  ...                       &      ...                        &  $33.82 \pm 0.25$       &        $2.17\pm 0.23$        &  ...    &  $<13$    &   ...         & \cite{2008AnA...477..353A}\\
41       &  $\!\!\!$[0.3]  &    $31.38\pm0.20$ &      $1.1\pm0.6$        &   ...                            &     ...                             &
                                                                                       
                                                                                             0.1  &  ...        &  ...          &  \cite{2004ApJ...612..398H} \\
42      &     2.0   &    $31.93  \pm   0.13$    &     $2.1\pm1.0$         &   ...                             &     ...                             &  0.4  &   ...       &     ...    &  \cite{2002ApJ...579..404P}    \\
43       &   0.3      &    $31.54  \pm 0.30$       &      $1.5\pm0.6$         &  ...                             &    ...                             &  0.3  &   ...        &      ...    & ... \\
44       &    $\lesssim1$  &  ...                      &     ...                          &  $32.00\pm0.50$        &    $[2.6]$                         & ...    &   ...        &    $<40$  & \cite{2009ApJ...700L.127A} \\
45      & $\!\!\!$[1.1] &    $31.60   \pm   0.50$  &      ...                &   $34.30   \pm 0.11 $     &    $2.31\pm0.23$           &  0.2  &   $ 50   $ &   $ 30$    &  \cite{ 2006ApJ...636..777A}  \\
46      & $\!\!\!$[0.3] &    $31.20\pm 0.50$       &      ...                 &   ...                              &     ...                            &  0.05 &   ...        &   ...    &  ... \\
47       & $\!\!\!$[1.5]         &    $31.53\pm 0.10 $      &   [1.5]                 &   $32.97\pm0.09$         &   $1.9\pm0.3$               & 0.5   &   6         &   4     &  \cite{2009ApJ...705....1C, 2002AnA...393L..37A,2007ApnSS.309...29M}  \\  
48      & $\!\!\!$[0.1] &    $31.11  \pm   0.10$    & $1.5\pm0.3$         &  ...                             &  ...                              &  0.8  &  ... &   ...    &  \cite{ 2008ApJ...684..542K}  \\

\hline
\end{tabular}
\caption{
Properties of the X-ray/TeV PWNe 
listed in Table 1}
\label{tab:prop2}
\end{table}

\addtocounter{table}{-1}
\begin{table}
 \setlength{\tabcolsep}{0.05in}
\begin{tabular}{rlcccclllc}
\hline
 \tablehead{1}{c}{b}{   \# \\  }  &
  \tablehead{1}{c}{b}{   $N_{\rm H,22}$
  \tablenote{ Hydrogen column density (in units of $10^{22}$ cm$^{-2}$)
obtained from spectral fits to the PWN spectra or
 estimated  from the pulsar's dispersion measure assuming 10\% interstellar medium ionization (in square brackets for the latter case). In a few cases upper limits are given based on the galactic HI column density.
  } \\  }  &
   \tablehead{1}{c}{b}{   $\log L_{\rm X}$\tablenote{Logarithm of
 PWN luminosity in the 0.5--8 keV band. The quoted errors are purely statistical, they do not include the distance errors. 
For
bright PWNe
(e.g., \#\#  2, 8, 16), we quote
the luminosity of the PWN ``core''
restricted to the torus/arcs
regions.
For the PWNe with extended tails
 (\#\#  39, 42, 48, 53, 54, 55, 58)
we quote only the luminosity of the bright ``bullet'' component. 
For \#\#  36, 45 and 46, faint extended emission is seen around the pulsar but its luminosity is
very uncertain; we use $\pm 0.50$ dex as a conservative estimate for the uncertainty.  
} \\ ${\rm [erg/s]}$ }  &
   \tablehead{1}{c}{b}{   $\Gamma_{\rm X}$  \\  }  &
   \tablehead{1}{c}{b}{   $\log L_\gamma$\tablenote{Logarithm of PWN (or candidate) luminosity
in the 1--10 TeV band. 
} \\ ${\rm [erg/s]}$  }  &
    \tablehead{1}{c}{b}{   $\Gamma_{\gamma}$\tablenote{Photon index of TeV spectrum determined from a power-law model. The fits are not good (e.g., an exponential cutoff is required or the spectral slope is nonuniform) in the cases marked by the superscript $^?$. The values in square brackets were assumed for estimating $L_\gamma$.}  \\  }  &
      \tablehead{1}{c}{b}{   $l_{X}$\tablenote{ Size of X-ray PWN ``core''
in which the PWN X-ray properties listed in this table were measured.}  \\  ${\rm pc}$ }  &
  \tablehead{1}{c}{b}{   $l_{\gamma}$\tablenote{
 Size of TeV source. If the source is unresolved, we quote the upper limit on $l_\gamma$.
}  \\ ${\rm pc}$ }  &
 \tablehead{1}{c}{b}{   $\Delta$\tablenote{
 Offset between the X-ray and TeV components.}  \\ ${\rm arcmin}$ } &
\tablehead{1}{c}{b}{ ${\rm Refs.}$\tablenote{
The PWN/PSR X-ray properties listed here were measured by ourselves
(except for \#\# 2, 5, 6, 10, 37, 47, 49 and 51), but
we cite recent relevant papers when available.  \vspace{-0.8cm} }\\}
  \\
\hline
49     &  $\lesssim0.6$ &   $<31.15$                         &   ...                       &     $32.64\pm0.50$     &   $[2.6]$                       & ...    &  $<180$   & ... &  \cite{2009ApJ...700L.127A, 2009arXiv0911.3063T}\\  
50      &     0.1   &    $30.81  \pm   0.18$    &     $1.6\pm0.5$         &  ...                               &    ...                              &  0.1  &   ... &  ...    &  \cite{2003Sci...299.1372S}    \\
51     &  $\lesssim0.04$ &        ...             &      ...                       &  ...                                &   ...                              &  1.5     &   ... &  ...    &   \cite{2009_Kawai_Fermi} \\ 
52         &   $\lesssim1.5$ &         ...             &     ...                        &   $32.82\pm0.15$          &   $2.78\pm0.31$           & ...     &  22   & 12  & \cite{ 2009arXiv0907.0574T} \\
53     &     0.25  &    $31.04  \pm   0.10$    &     $3.3\pm0.5$         &  ...                               &   ...                              &  0.2 &           ... & ...   &  \cite{2003ApJ...585L..41R, 2007ApJ...654..487N} \\
54      &     0.6   &    $31.19  \pm   0.03$    &     $1.5\pm0.1$         &   ...                               &   ...                              &  0.1 &   ...          &  ...   &  \cite{2006ApJ...647.1300M}    \\
55      &     0.03  &    $29.71  \pm   0.07$    &     $1.0\pm0.2$         &   $30.22\pm0.50$         &      $[2.6]$                    &  0.02 &   $10$ &  $<30$    &  \cite{2006ApJ...643.1146P, 2009ApJ...700L.127A}    \\
56       &    $\lesssim1.3$ & $<31$              &       ...                       &    $\sim33.01$             &   $1.82\pm0.35$            &  ...    &   $6$ &  $7$        &  \cite{2009arXiv0907.0574T}   \\  
57       &     $\lesssim2$ &            ...           &              ...                &   $34.66\pm0.60$          &   $2.4\pm0.44$              & ... &     $70$  &  ...          &   \cite{ 2009arXiv0907.0574T}  \\
58      &     0.17  &    $29.63  \pm   0.01$    &     $1.7\pm0.6$         &   ...                              &   ...                             &  0.05 &   ...    &  ...  &  \cite{ 2008ApJ...685.1129M} \\
59      &     0.2   &            $29.5\pm0.5$     &         $[1.5]$             &   ...                             &  ...                              &  0.07  &     ...     &   $   ...  $    &  \cite{2007AandA...467.1209H}    \\
\hline
\end{tabular}
\caption{
Properties of the X-ray/TeV PWNe 
 listed in Table 1}
\label{tab:prop2}
\end{table}

As of  December 2009,
 about 60 
PWNe associated with known radio or $\gamma$-ray pulsars have been detected.
  We have compiled the properties of these  PWNe and their host pulsars in Tables 1 and 2.  In most cases we have reanalyzed the X-ray PWN
  properties 
 to ensure the uniformity of the analysis. 
The luminosities, spectral slopes, and sizes of the TeV  PWNe have been taken from  publications as well as from recent conference presentations. 
 Because of the 
 page limit, we  mainly discuss the luminosities and, very briefly, spectra and sizes of these PWNe.  
 
 We see from Table 2 that the 
 TeV PWN sizes generally  increase with pulsar age  while the X-ray PWN sizes\footnote{ Here,
 while referring to X-ray PWN size, spectrum, or luminosity, we only consider the bright PWN ``core''
  restricted  to the torus/arcs regions or the ``bullet''  in ram pressure confined PWNe (see \cite{2008AIPC..983..171K}).}
 show an opposite trend.  
 Moreover, for pulsars older
        than $\sim 10$ kyr the sizes of the TeV PWNe
        are typically 100--1000 times larger than the sizes of the  X-ray PWNe, while the difference is only a factor of a few
        for some younger pulsars (e.g., Crab). 
 This suggests that 
  the aged and cooled electrons (mainly seen through their synchrotron emission in radio and  IC emission in 
 TeV) have 
 propagated, 
 through advection and/or diffusion, farther away from the pulsar 
than the recently 
  injected electrons, responsible for the X-ray synchrotron nebulae.  
While comparing the X-ray and TeV PWNe, one should take into account that
larger numbers of electrons have been produced earlier in the pulsar's life 
 (the electron production rate
$\propto \dot{E}(t) \propto \dot{E}_0 (1+t/\tau_0)^{-(n+1)/(n-1)}$,
 where $n$ is the pulsar's braking index and $\tau_0$ is the initial spin-down timescale;
e.g., \cite{2006ARA&A..44...17G}). 
This picture is further complicated by 
environmental effects and pulsar motion.
For instance, the 
 ambient pressure 
 that confines a slowly-moving  PWN,  
is much greater in a star-forming region than in low-density regions
 above the Galactic plane; 
high-speed pulsars are accompanied by long tails 
   with higher bulk flow velocities compared to more isotropic PWNe around slowly-moving pulsars; 
  pressure gradients inside a supernova remnant  or interaction with its reverse shock can 
  affect the PWN shape.
        In addition, 
    the  ambient radio/IR radiation density may vary and  affect the properties of the IC radiation emitted by the pulsar wind.  
In many cases, significant offsets  between the centroid of the 
        extended TeV source and the neighboring pulsar's position 
 may be attributed to the interaction 
with the asymmetric reverse shock or to a locally non-uniform distribution of the ambient matter/radiation. 
      The   large sizes of these TeV sources and the large offsets 
         may sometimes lead to false associations 
 with pulsars, 
 especially if there is no preferential extension of the X-ray PWN toward the TeV source center. 

The X-ray and TeV spectra of  PWNe 
are often approximated by 
 power-law (PL) models with different slopes. 
If the electron spectral energy distribution (SED) can be 
 approximated by a single PL, $dN(\epsilon)\propto \epsilon^{-p}d\epsilon$, 
then both the synchrotron (X-ray) and  IC (TeV) spectra should 
 have the same slopes, $F_{\nu}\propto\nu^{-\Gamma+1}$, where the photon index $\Gamma=(p+1)/2$. 
However, one can see from Table 2  that the spectra of X-ray PWNe are systematically harder than those 
of their TeV counterparts, 
which can be attributed to the evolution of the electron SED.
    Indeed, \chan\ observations of  bright, well-resolved PWNe have shown that 
the synchrotron  spectrum varies significantly with the distance from the pulsar 
(due to the radiation and expansion energy losses),
 and therefore a single PL approximation to the X-ray spectrum becomes 
 inapplicable if a large 
 enough volume around the pulsar is considered. Also, the IC spectrum may deviate from a PL at large energies due to 
 the intrinsic  upper boundary of the electron SED, or due to the cooling effects.  In addition to the spectral differences 
 caused by the cooling and boundary effects, one should keep in mind that while the intensity of the synchrotron component depends on the magnetic field strength,
 the IC intensity depends on the ambient radiation density instead.
    Therefore, there is no reason to expect direct 
 proportionality between the X-ray and TeV fluxes of a particular PWN, and indeed we see 
   little correlation 
 between the X-ray and TeV luminosities (Fig.~1, panel A). Yet, the X-rays, observed from the innermost PWN region, 
and TeV $\gamma$-rays, produced throughout the entire PWN volume, 
   can be related 
 in a model that takes into account the pulsar wind history (e.g., the evolution of $\edot$
 and the PWN size), 
its spatially varying properties,
  and local radiation density.  
 While measured from spatially different regions of a PWN, the X-ray and TeV luminosities in Table 2 still can be used to test such models 
as long as the observed values are compared with the model values calculated for appropriate PWN zones 
 and time intervals.
 There have been a number of analytical models that take into account either 
the spatial dependence (e.g., \cite{1984ApJ...283..694K,1984ApJ...283..710K}) or time-dependence (e.g., \cite{2000ApJ...539L..45C, 2009ApJ...703.2051G}) of the  PWN properties but rarely both (\cite{2000AnA...359.1107A}).

 \begin{figure}
 \centering
\includegraphics[width=1.0\textwidth,angle=0]{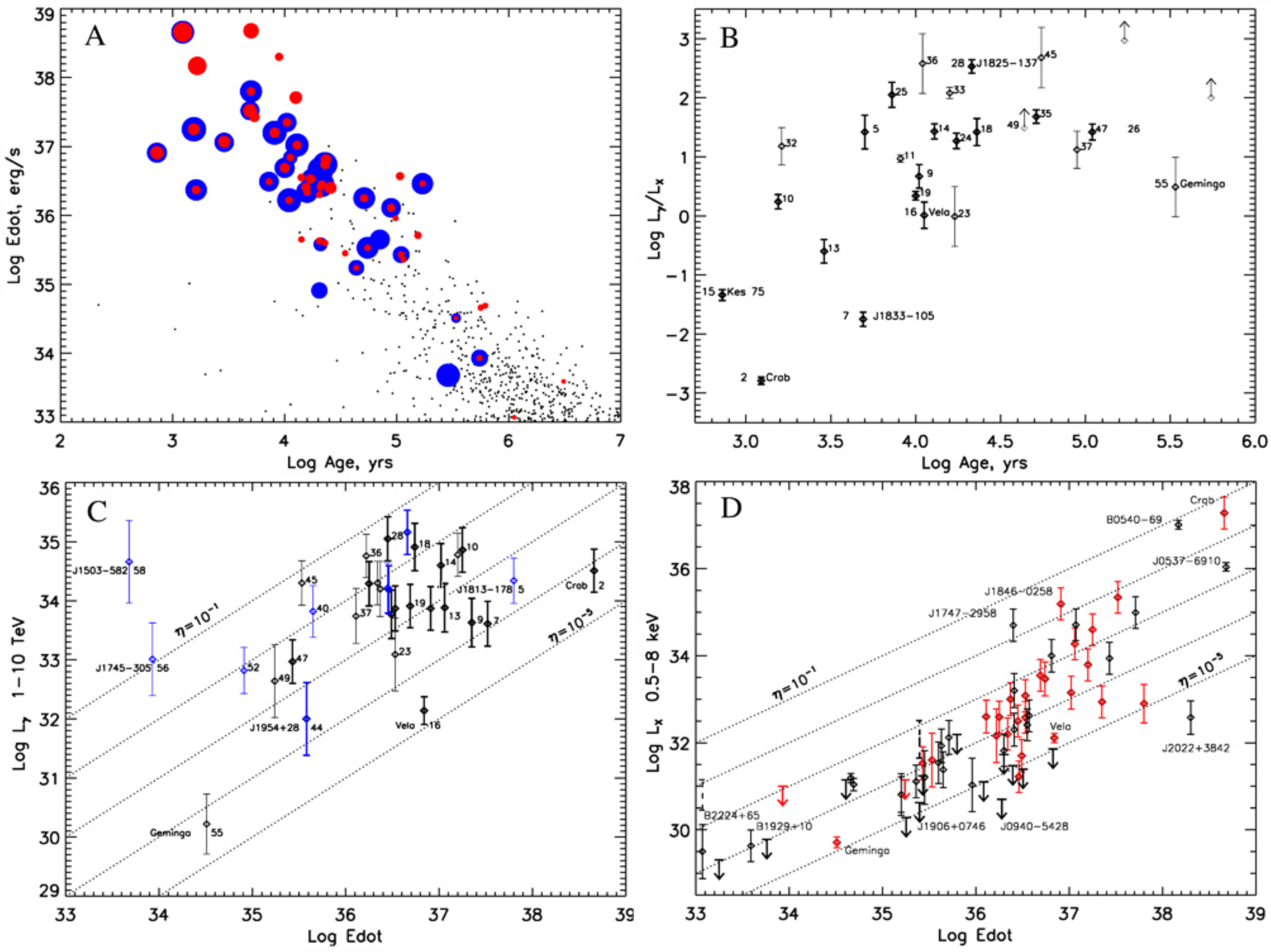}
\caption{    {\em (A)}  Filled circles are 
X-ray (red) and TeV (blue) detected PWNe or PWN candidates. 
 Larger circle  sizes correspond to higher luminosities.
The small black 
 dots denote the pulsars from the ATNF catalog (\cite{2005AJ....129.1993M}).   {\em (B) }  Ratio of the TeV to X-ray luminosities
   vs.\ pulsar's spin-down age. Here and in panel {\em C} the thin error bars mark questionable associations.  {\em (C) }   TeV luminosities of PWNe and PWN candidates
 vs.\  pulsar's $\dot{E}$.
 The blue error bars mark TeV objects without X-ray counterparts.
Here and in panel {\em D} the luminosity uncertainties include 50\% systematic uncertainty assigned to the  distances
unless the pulsar is in LMC or its parallax has been measured. {\em (D)} X-ray luminosities of PWNe  vs.\  $\dot{E}$.  The red error bars denote X-ray PWNe with
  TeV counterparts.  The lines of constant 
 radiative efficiency ($\eta\equiv L_{X,\gamma}/\dot{E}$) are plotted in panels  {\em C}  and  {\em D}.
\vspace{-0.8cm} }
\end{figure}

In  panels C and D of Figure 1 we  
show the dependences of the luminosities $L_{X}$ and $L_{\gamma}$ on 
 $\edot$.
 As expected, we 
  do not see an obvious correlation between $L_\gamma$ and $\dot{E}$, 
 which likely reflects the fact that  the TeV luminosity depends on the history of the pulsar spin-down rather than on the current $\dot{E}$. 
   An additional 
 scatter is expected due to  the differences in the local IR background and the uncertain distances. Most of the 1--10 TeV luminosities cluster 
in the range of  $10^{33}$--$10^{35}$ erg s$^{-1}$, a much narrower span than that attained by the X-ray luminosities shown in 
 another panel.
 The $L_{X}$ versus $\edot$  plot does show 
some correlation between the two; however, the  more 
 objects are included the  larger  the scatter becomes,
weakening the correlation.   
It seems very  unlikely that  
 a factor of $10^{4}$ 
 scatter at a given
 $\edot$ could be explained by varying environment (e.g., ambient pressure) 
or differences in pulsar velocities. 
 Some ``hidden'' pulsar parameters (e.g., the angle between the rotation and magnetic axis, or the topology of the NS magnetic field)  
may instead govern the efficiency of the magnetic-to-kinetic energy conversion and, as a result, the X-ray efficiencies of PWNe 
(see also \cite{2008AIPC..983..171K}). 
  In Figure 1 (panel B) we also plot the distance-independent  
$L_{\gamma}/L_X$ as a function of spin-down age $\tau$, which shows
 a  hint of positive  correlation between the two, at younger ages.
 This correlation can be explained by 
 the correlation between $L_X$ and $\edot$ (hence 
the anti-correlation between $L_X$ and $\tau$). 
These results are in general agreement with the recent findings of \cite{2009ApJ...694...12M} 
 based on a smaller sample of PWNe.  At $\log\tau\gtrsim4.5$  the dependence of $L_{\gamma}/L_X$ on $\tau$ appears to level off as expected from the simple model of 
  \cite{2009ApJ...694...12M}.  
 This trend, however, shows a large scatter,
 which might be attributed to the magnetic and radiation fields being different for different objects.

   To conclude, multiwavelength observations of PWNe are crucial because they provide identifications for VHE sources and reveal the true energetics and 
   composition of pulsar winds.  In addition to  X-ray and TeV observations, 
 there is an opportunity  to detect the IC PWN component with {\sl Fermi } in GeV, 
    where even old objects are expected to exhibit uncooled 
     IC spectrum  which should  be more directly linked to the radio synchrotron component. 
The data accumulation must be complemented by 
 development of multi-zone models of PWN evolution,  essential for understanding the role 
   of pulsar winds in seeding the Galaxy with energetic particles and magnetic fields.   


\begin{theacknowledgments} 
 We would like to thank Wenwu Tian for providing updated distances for several sources.  
   The authors acknowledge support from NASA 
 grants NNX09AC81G  and
   NNX09AC84G,    and {\sl Chandra}
   award   AR8-9009X.  This material is also based upon work supported by the National Science Foundation under Grants No. 0908733 and 0908611.
   
 \end{theacknowledgments}



\bibliographystyle{aipprocl} 

\bibliography{astroph.bib}

\IfFileExists{\jobname.bbl}{}
 {\typeout{}
  \typeout{******************************************}
  \typeout{** Please run "bibtex \jobname" to optain}
  \typeout{** the bibliography and then re-run LaTeX}
  \typeout{** twice to fix the references!}
  \typeout{******************************************}
  \typeout{}
 }

\end{document}